\documentstyle{article}
\newcommand{\bc}{\begin{center}}
\newcommand{\ec}{\end{center}}

\newcommand{\barr}{\begin{array}}
\newcommand{\bey}{\begin{eqnarray}}
\newcommand{\be}{\begin{equation}}
\newcommand{\ear}{\end{array}}
\newcommand{\eey}{\end{eqnarray}}
\newcommand{\ee}{\end{equation}}

\newcommand{\pde}{\partial}

\newcommand{\spao}[1]{\mbox{\hspace{#1}}}
\newcommand{\spav}[1]{\parbox{1mm}{\vspace*{#1}}}

\newcommand{\ssty}{\scriptstyle}
\newcommand{\sssty}{\scriptscriptstyle}

\newcommand{\text}{\textstyle} 

\newsavebox{\ipiu}
\newsavebox{\imen}

\renewcommand{\a}{\alpha}
\renewcommand{\b}{\beta}

\renewcommand{\d}{\delta}

\renewcommand{\l}{\lambda}
\renewcommand{\L}{\Lambda}

\sbox{\ipiu}{$\ssty i \sssty +1$}
\sbox{\imen}{$\ssty i \sssty -1$}

\begin{document}


\rightline{DPS-NA-18/95}
\rightline{INFN-NA-18/95}
\rightline{DFUPG-101/95}
\bc \spav{1cm}\\
{\LARGE\bf Abelian Hall Fluids and Edge States:\\
a Conformal Field Theory Approach}

\bigskip

{\large A. De Martino \\
{\normalsize\em  GNSM Sez. di Perugia\\}
{\normalsize\em c/o Dipartimento di Fisica, Universit\`a di Perugia\\}

 and \\ R. Musto  \\}
{\normalsize\em Dipartimento di Scienze Fisiche, Universit\`a di Napoli\\}
{\normalsize\em and INFN Sez. di Napoli\\}
\spav{1.6cm}\\
{\small\bf Abstract\\}
\spav{2mm}\\
{\small\parbox{13cm}{\spao{4mm}
We show that  a Coulomb gas Vertex Operator representation of 2D
Conformal Field Theory gives a complete description of abelian Hall
fluids: as an euclidean theory in two space dimensions leads to the
construction of the ground
state wave function for planar and toroidal geometry and characterizes the
spectrum of low energy
excitations; as a $1+1$ Minkowski theory gives the corresponding dynamics of
the edge states. The difference between a generic Hall fluid and states of
the Jain's sequences is emphasized and the presence, in the latter case, of
of an
$\hat {U}(1)\otimes \hat {SU}(n)$ extended algebra and the
consequent
propagation on the edges of a single charged mode and $n-1$ neutral modes
is discussed.
}\\}
\spav{4mm}\\

\bigskip
\bigskip
\bigskip
\baselineskip=12pt

\bigskip
\bigskip
\ec
\newpage

\large

{\bf 1. Introduction}

Since the fractional quantized Hall effect (FQHE) was first observed by
Tsui, Stormer and Gossard$^1$ in 1982, considerable experimental
and theoretical progress$^2$ has been made toward a physical
understanding and a
formal
characterization  of this intrinsically collective
phenomenon. An important element, common to many recent
developments, is  a better understanding of the interplay
between the physics of the
incompressible Hall fluid of the bulk and the dynamics of the gapless
excitations on the edges of the sample. Indeed, for the simple case of
filling $\nu =1$,
the quantization of the
Hall conductance may be related to the perfect transmission of free
electrons edge states$^3$ and, more generally, even in the
case of multiple edge channels, stability of the edge currents
against impurity perturbation is expected to be crucial for explaining the
experimentally observed Hall plateaux.

A natural and unified description of the  properties of the bulk and of the
edges of a Hall system is provided by a
2D Conformal Field Theory (2DCFT) approach$^4$. In this
framework the main tool is a Coulomb gas Vertex Operator
representation, that translates in
a precise mathematical form the physical idea that the quasiparticles of the
Hall fluid arise from the binding of electrons
and magnetic vortices. In the case of the Laughlin states$^5$,
corresponding to a filling factor $\nu =1/m$, the electron
Vertex Operator (VO) is a $U(1)$ field of conformal weight $m/2$ and the
corresponding ground state wave function (gswf) is given by an appropriate
correlator of such fields. Therefore the Laughlin wave function satisfies a
Knizhnik-Zamolodchikov equation for an abelian Wess-Zumino
model$^6$, expressing the existence of a non trivial
connection due to the presence of infinitely
thin magnetic flux tubes located at the positions of the  particles.
Furthermore, one
can show that in order to build the Laughlin gswf's on a compact
Riemann surface  one needs a set of
VO's, characterized by the integer lattice
$\text{{\bf Z}} / m\text{{\bf Z}}$, in one-to-one
correspondence with the $m$-fold degeneracy of the
gswf's on a torus,  leading  to a field theoretical
characterization of topological order
and providing the spectrum of low lying excitations.

It has been stressed$^{7,8}$ that FQHE on closed Riemann surfaces, although
experimentally inaccessible, dictates the structure of the edge dynamics.
The same $U(1)$ 2DCFT, that as an euclidean theory in two space dimensions
leads to
the construction of the gswf's, gives, as a theory  in $1+1$ space-time
dimensions,
the dynamics of the
chiral edge states.
Indeed, by looking at the edge states on a cylinder, one recognizes
the existence of $m$ sectors of fractionally charged excitations,
that are described by means of the same set of $m$ VO's required to build
the gswf's on a torus, expressed  in this case in terms
of a chiral field propagating on the edge.

The purpose of this paper is to  extend this picture in the general case of
an abelian quantum Hall fluid, that can be characterized$^9$
by a symmetric integer
valued $n
\times n$
matrix, $K$. We shall  see  that in order to have well defined gswf's
on the torus the $K$ matrix must be positive definite. We may
recall that
this requirement implies that the multiple channel edge currents
move all in the same
direction$^{10}$,  and, as a consequence,
the total current is not altered by scattering events
and the  conductance is quantized. Furthermore it has been shown$^{11,12}$,
for Jain's states$^{13}$ corresponding to $K$ matrix not positive
definite, that it is
necessary to take into account the effect of disorder
to explain the  observed Hall conductance
quantization. Therefore in the
following we shall assume that the positivity  condition for the $K$ matrix
is fulfilled and we shall show that
a complete 2DCFT description of
a multi component abelian Hall fluid can be achieved by introducing
$n$ properly compactified, holomorphic scalar fields.
The complete set of inequivalent
VO's will be characterized  by the $n$
dimensional integer lattice $\text{{\bf Z}} ^n /K \text{{\bf Z}} ^n$,
whose points
are in one-to-one correspondence with the $\det K$ degenerate gswf's on the
torus. This set of VO's will also lead to a full
characterization of the low energy fractionally
charged excitations and the corresponding edge states.

However there are important differences in the structure of the
2DCFT between the case of a {\it generic}- in a sense to be
later specified- $K$  matrix and the case of the matrix corresponding to
a Jain's state of filling $\nu= n/(2np+1)$.
We shall see that in the latter case the lattice characterizing the
inequivalent VO's may be taken as $\text{{\bf Z}}/(2np+1)\text{{\bf Z}}$,
 in complete analogy with the
Laughlin states. Furthermore, an extended algebra
$\hat {U}(1)\otimes \hat {SU}(n)$ will naturally appear and,  by introducing
a formal description of the transport mechanism,
we shall see that only the
$\hat {U} (1)$ mode contributes to the conductance, while the ones
corresponding to
$\hat {SU} (n)$ are neutral.
The important consequence for the dynamics
of the edge states is that, while in the generic case there will be $n$
charged modes propagating on the edges, in the Jain's case there will be a
single charged mode and $n-1$ neutral ones. This  picture, that confirms the
results obtained in refs. 12,14, may
be a starting point for understanding the experimental evidence showing that
the most prominent Hall plateaux belong to the Jain's sequences.

This paper is organized as follows. In sec. {\bf 2} we briefly review the
results relative to the Laughlin states, stressing the correspondence between
the 2DCFT description on the bulk and on the edge. Sec. {\bf 3} is
devoted to a general abelian Hall fluid. Once a complete set
of VO's is identified, the corresponding gswf's on the torus
are constructed, their
properties under modular transformations and magnetic translations
are analyzed and the response to an applied electric field
is evaluated.
In sec. {\bf 4} the corresponding
description for the edge states is given. The Jain's states are discussed
in sec. {\bf 5}, where it is shown that in a basis that diagonalizes $K$
an $\hat {U}(1)\otimes \hat {SU}(n)$ structure
naturally arises; only the
$\hat {U}(1)$ mode contributes to the conductance and the center
of charge theta function can be decomposed as a sum of  products of factors
relative to the charged and the neutral modes. Taking the $\hat {SU}(2)$ and
$\hat {SU}(3)$
cases as specific examples, the corresponding
character decomposition is exhibited.
Finally, sec. {\bf 6} is devoted to concluding
remarks and perspectives.

{\bf 2. Laughlin sequence}

In order to present the basic ideas relative to the description of the
quantum Hall
effect by means of 2DCFT and fix the basic notations, we recall briefly the
well known results for the Laughlin sequence$^4$.
The starting point is the introduction of a
 holomorphic scalar field $\phi (z)$,
with two point correlator:
\begin{equation}
\label{1}
\left\langle \phi \left( z_1\right) \phi \left( z_2\right)
\right\rangle =-\ln \left( z_1-z_2\right) \, .
\end{equation}
The mode expansion for the field $\phi \left( z\right) $
is given by
\begin{equation}
\label{2}\phi \left( z\right) =\hat {q}-i\hat {p}\ln z+i\sum_{n\neq
0}\frac{a_n}%
nz^{-n} ,
\end{equation}
with coefficients satisfying the usual commutation relations.
By taking the field $\phi (z)$ compactified on a circle of radius
$R=\sqrt{m}$, i.e. $\phi \equiv \phi +2\pi \sqrt  m$, one identifies a set of
$m$ inequivalent VO's:
\begin{equation}
\label{4}
V_l \left( z\right) =:\exp \left[ i{l\over \sqrt m }\phi \left(
z\right) \right]:=:\exp \left[ il\sqrt{\nu }\phi \left(
z\right) \right]:\, , \quad \quad \quad l\in \frac{\text{{\bf Z}}}{m
\text{{\bf Z}}}
\, .
\end{equation}
The operators $V_l$ are primary fields of conformal weight $\Delta
=l^2/2m$. Their physical meaning is easily recognized  by looking
at the braiding relation:
\begin{equation}
\label{5}
V_{l}\left( z\right) V_{l'}\left( z'\right)
=\left( z-z'\right) ^{l l' \over m}:V_{l }\left( z\right)
V_{l'}\left( z'\right) :=e^{i\pi {l l' \over m}} V_{l' }\left( z' \right)
V_{l }\left( z\right)\, ,
\end{equation}
showing that the VO $V_l$ describes a particle with statistical factor
$l^2/2m$,
carrying $l$ units
of magnetic flux and
electric charge $l/m$.
The electron field corresponds to the choice $l=m$ and the analytic
part of the Laughlin gswf is given by
\begin{equation}
\label{6}\left\langle V_{\sqrt{m}}\left( z_1\right) \dots V_{\sqrt{m}%
}\left( z_N\right) \right\rangle =\prod_{i<j}^N\left( z_i-z_j\right) ^m,
\end{equation}
where by definition
\begin{equation}
\label{7}\left\langle \prod_{i=1}^N V_{\sqrt m}\left( z_i\right)
\right\rangle \equiv \left\langle N{\sqrt m}\left| \prod_{i=1}^N
V_{\sqrt m
}\left( z_i\right) \right| 0\right\rangle
\end{equation}
and
\begin{equation}
\label{8}
\left\langle p \right |\hat{p} =p \left \langle p \right |,
\quad \left \langle p \right | a_{-n}=0\, , \quad  n>0\, .
\end{equation}

Due to the finite energy gap existing for an incompressible Hall fluid, the
full spectrum of low energy excitations, corresponding to the full set of
operators given by eq. (\ref{4}), will not play a
crucial role at very low temperature.
However, their theoretical relevance is due to the one-to-one
correspondence between the set of
VO's eq. (\ref{4}) and the set of vacuum states $\left| g=1,\, l
\right\rangle$,
that enter in the construction of the Laughlin gswf's on a compact
genus one Riemann surface:
\begin{equation}
\label{9}\left\langle \prod_{i=1}^N V_{\sqrt{m}
}\left( z_i\right)
\right\rangle _l^{g=1}\equiv \left\langle N{\sqrt {m}}\left|
\prod_{i=1}^N V_{\sqrt{m}} \left( z_i\right) \right| g=1,l\right\rangle
\end{equation}

More explicitly, on a torus described by $z/L=\xi+\tau \eta$, Im$\tau>0$, with
$\xi \equiv \xi +1$ and $\eta \equiv \eta +1$,
 pearced
by an integer number $N_{\Phi}$ of magnetic fluxes,
in the Landau gauge $\vec A=2\pi N_{\Phi}(\eta,0)$, one has
\begin{equation}
\label{10}
\left\langle \prod_{i=1}^{N} V_{\sqrt{m}}\left(
z_i\right) \right\rangle _l^{g=1}=
\prod_{i<j}^{N}\left[
\theta _1\left(
 z_{ij}|\tau \right)
\right] ^m
\Theta \left[
\begin{array}{c}
l/m \\
0
\end{array}
\right] \left( mZ|m\tau \right)
\equiv F_l  \left( \left\{z_i\right\}|\tau\right)\, ,
\end{equation}
where $Z=\sum_{i=1}^{N} z_i/L$ is the center of charge variable and
$z_{ij}=(z_i - z_j)/L$.
Here we have introduced the theta functions with rational
characteristics$^{15}$
defined by
\bey
\label{11}
\Theta \left[
\begin{array}{c}
a \\
b
\end{array}
\right] \left( z|\tau \right) &=&\sum_{h\in \text{{\bf Z}} }\exp \left[ \pi
i\left( h+a\right) ^2\tau +2\pi i\left( h+a\right) \left( z+b\right)
\right]=
\nonumber
\\
&=&{\cal S}_b {\cal T}_a \Theta \left[
\begin{array}{c}
0 \\
0
\end{array}
\right] \left( z|\tau \right)\, .
\eey
where $a,b \in \text{{\bf Q}}$.

The above equation defines implicitly the magnetic translation operators
$\cal S$ and $\cal T $. The presence of the product of theta functions
of the relative particle positions in eq. (\ref{10}) is expected because
$[-\ln\,\theta_1(z_i-z_j|\tau)]$ is the singular part of the Coulomb propagator
on the torus as $[-\ln(z_i -z_j)] $ is
on the plane. On the other hand, the center of charge theta function
is required in order
to satisfy the correct boundary conditions in each particle
variable and to give a uniform charge distribution in the thermodynamical
limit, as can be easily shown provided the condition $N_{\Phi} =m N$,
expressing the cancellation (in the average) between
the external magnetic field and the statistical one, is fulfilled.

The gswf's eq. (\ref{10}) realize a natural splitting between local and
global properties: while the local part is unaffected by total magnetic
translations, the center of charge
theta functions, and henceforth the complete gswf's, provide an irreducible
representation of the subgroup of the total magnetic translation
group generated by ${\cal S}_{L/N_{\Phi}}$ and ${\cal T}_{L/N_{\Phi}}$:
\be
\label{12}
\begin{array}{l}
{\cal S}_{L \over N_\Phi }F_l=\exp \left[ 2\pi i\frac lm\right] F_l \, ,
\\
{\cal T}_{L \over N_\Phi }F_l=F_{l+1}\, .
\end{array}
\ee
The Hall conductance quantization
can be seen as a global property:
by introducing, along
the $B$ cycle of the torus, a magnetic flux tube
of strength $2\pi \lambda $  the Laughin gswf's
are transformed as follows:
\be
\label{13}
F_l \rightarrow  F_{ l + \l }={\cal T}_{\lambda L \over N_\Phi}F_l \, ,
\ee
implying$^4$ a pure Hall conductance $\sigma_H = 1/m$ in natural units.

Although the Hall effect on closed Riemann surfaces is not experimentally
accessible, the above argument has immediate consequence on the
structure of the edge state excitations.
Indeed, consider the cylinder resulting from
cutting the torus along the
$A$ cycle; then, from eq. (\ref{13}), one sees that the variation of
one unit of flux
implies the transfer of a charge $1/m$ from one edge to the other. The charge
spectrum of edge states excitations is then in one-to-one correspondence with
the set of VO's eq. (\ref{4}). Furthermore, the same VO's
give  {\it bosonized} field operators representing the charged
edge excitations,
where now $\phi =\phi (x-vt)$ is a chiral field  in $1+1$
space-time, and $x$ is a coordinate along the edge.

The 2DCFT provides a complete description of
the dynamics of the edge states
as the Hamiltonian density of the neutral excitations
and the electric current
are completely expressed in terms of the chiral field $\phi = \phi (x-vt)$:
\begin{equation}
\label{14}
{\cal H}=\frac v{8\pi }\left[ \frac 1{v^2}\left( \partial
_0\phi \right) ^2+\left( \partial _x\phi \right) ^2\right] \, ,
\end{equation}
\begin{equation}
\label{15}j=\frac{\sqrt{\nu}}{2\pi } \partial _x\phi \,  ,
\end{equation}
The current satisfies a $\hat {U}(1)$ Kac-Moody algebra that
in terms of Fourier components reads
\begin{equation}
\label{16}\left[ j_n,j_m\right] ={\nu}n\delta _{n+m,0} \, ,
\end{equation}
and the following commutation relation with the Hamiltonian
\begin{equation}
\label{17}\left[ H,j_n\right] =-vnj_n \, .
\end{equation}
In order to stress the difference with an ordinary Fermi
liquid, Wen$^7$ has introduced the term {\it chiral Luttinger liquid} to
describe such a system.

{\bf 3. Abelian Hall Fluid}

A general $n$ component Hall fluid may be
characterized$^9$ by a symmetric integer valued matrix $K$, where the element
 $K_{IJ}$ is the braiding factor between an electron of the $I^{th}$
component and an electron of $J^{th}$ component.
To describe such a system in the framework of 2DCFT
we introduce a set of $n$ independent holomorphic scalar fields ${\phi}_i (z)$
whit correlators
\begin{equation}
\label{19}
\left\langle \phi _i\left( z_1\right) \phi _j\left( z_2\right)
\right\rangle =-\delta _{ij}\ln \left( z_1-z_2\right) \, .
\end{equation}
The correct braiding properties between any two electrons are obtained
by defining the VO's
\begin{equation}
\label{20}
V_{\vec {\b _I}} \left( z\right) =:\exp \left[ i \vec
{\b _I}
\cdot \vec
\phi \left( z\right) \right]: \, ,
\end{equation}
where the vectors $\vec \b _I$ satisfies $\vec \b _I \cdot \vec
\b _J = K_{IJ}$. A more explicit form for $\vec \b _I$ will
 be given
below. The conformal weight, and therefore the spin, of the VO's
eq. (\ref{20}) is given by ${\vec \b _I}^2/2$, leading to the
requirement $K_{II}$ odd. The explicit form of the generalized Laughlin
wave function on
the plane is given by
\begin{equation}
\label{21}
\left\langle \prod_{I=1}^n\prod_{i=1}^{N_I}V_{\vec \b _I
}\left( z_i^I\right) \right\rangle =
\prod_{I=1}^n\prod_{i<j}^{N_I}\left( z_i^I-z_j^I\right) ^{K_{II}}
\prod_{I<J}^n\prod_{i=1}^{N_I}\prod_{j=1}^{N_J}
\left( z_i^I-z_j^J\right)
^{K_{IJ}} \, .
\end{equation}
In order to obtain a complete characterization of the low energy excitations
of the system, we proceed as in the case of the Laughlin states and
introduce a full set of VO's. To this purpose,
we compactify the field $\vec \phi$ as follows: $\vec \phi \equiv
\vec\phi + 2\pi R\vec h $, where $\vec h \in \text{{
\bf Z}}^n$, and $R_T R=K$;
the explicit form of $R$ can be easily obtained by the diagonalization of $K$.
As a
consequence, a complete set of inequivalent VO's is given by
\begin{equation}
\label{22}
V_{\vec l}\left( z\right)=
:\exp \left[ i {\vec l}_T R^{-1}
\vec \phi \left( z\right) \right]: \,
, \quad \quad \quad \vec l\in \frac{%
\text{{\bf Z}}^n}{K\text{{\bf Z}}^n}\equiv \text{{\bf Z}}_K^n \, .
\end{equation}
Notice that the electronic VO's $V_{\vec \b _I}$ correspond to the choice
$\vec l=K\vec \d _I$, where $(\vec \d _I) _J=\d _{IJ}$.
The VO's eq. (\ref{22})
represent excitations currying a vector of magnetic flux $\vec l$ and
''charges''
\begin{equation}
\label{q}
Q_I=\left( K^{-1}\right) _{IJ}l_J\, ,
\end{equation}
as it can be seen from the braiding relation
\begin{equation}
\label{BR}V_{\vec l}\left( z\right) V_{\vec l^{\prime }}\left( z^{\prime
}\right) =\exp \left[ i\pi \vec l_TK^{-1}\vec l'\right] V_{\vec l^{\prime
}}\left( z^{\prime }\right) V_{\vec l}\left( z\right)\, .
\end{equation}
By evaluating the Hall conductance, we shall see shortly that the electrical
charge is given by $Q=\sum_IQ_I$. Notice that the electron charge is one as
it should.

The set of VO's eq. (\ref{22}) realizes a consistent description
of the system, as it can be seen analyzing
the Laughlin gswf's on the torus.
We work in the Landau
gauge $\vec A=2\pi N_{\Phi} (\eta,0)$ and we assume that
the condition
\be
\label{23}
K_{IJ}N_J=N_{\Phi} \, ,
\ee
expressing the
cancellation (in the average) between statistical and
external magnetic field, is fulfilled, implying that the system is at
filling factor $\nu =\sum_I N_I/N_{\Phi} =\sum_{I,J}\left( K^{-1}\right)
_{IJ}$.

Then the Laughlin gswf's on the torus
are given by
\be
\label{26}
F_{\vec l} \, \left ( \left\{ z_i^I \right\}|\tau \right )=\left\langle
\prod\limits_{I=1}^n\prod\limits_{i=1}^{N_I}V_{{\vec \b}_I}
\left( z_i^I\right) \right\rangle _{\vec
l}^{g=1},
\ee
where we have used again
the one-to-one correspondence existing in 2DCFT between
the set
of VO's (\ref{22}) and the $g=1$ vacuum states on the torus.
The explicit form of the gswf's is given by
\bey
\label{24}
F_{\vec l} \left( \left\{z_i^I\right\}|\tau \right)
&=&\prod\limits_{I=1}^n\prod\limits_{i<j}^{N_I}\left[ \theta_1
\left(
z_{ij}^{II}| \tau \right) \right] ^{K_
{II}
}\prod\limits_{I<J}^n\prod\limits_{i=1}^{N_I}\prod\limits_{j=1}^{N_J}%
\left[ \theta_1\left( z_{ij}^{IJ} | \tau \right) \right] ^{K_{IJ}}\times
\nonumber \\
&\times& \Theta \left[
\begin{array}{c}
K^{-1}\vec l \\
0
\end{array}
\right] \left( K\vec Z|K\tau \right) ,\quad \quad \quad
\,\vec l\in \text{{\bf Z}}_K^n \, ,
\eey
where $z_{ij}^{IJ}=(z_i^I -z_j^J)/L$, and
$ Z^I=\sum_{i=1}^{N_I} z_i^I/L$ is the center of charge coordinate of the
$I^{th}$
component. For the sake of simplicity, in the following we will use the
shorthand notation $\prod\nolimits^{\prime}
\left[ \theta_1 \right] $ for the
products of $\theta _1$-functions appearing in the right hand side of the
eq. (\ref{24}).
Here we also have introduced the theta functions of several variables
with rational characteristics$^{15}$:
\begin{equation}
\label{25}\Theta \left[
\begin{array}{c}
\vec a \\
\vec b
\end{array}
\right] \left( \vec z|\Omega \right) =\sum_{\vec h\in {\text{{\bf Z}}}^n}\exp
\left[ \pi i\left( \vec h+\vec a\right) _T\Omega \left( \vec h+\vec a\right)
+2\pi i\left( \vec h+\vec a\right)\cdot
\left( \vec z+\vec b\right) \right]\, ,
\end{equation}
where $\vec a,\vec b\in \text{{\bf Q}}^n$ and
$\Omega$ is a $n\times n$ symmetric complex valued matrix
with  positive definite imaginary part.
As $\text{{Im}} \tau >0 $ , it is now clear why we have taken
$K$ to be positive definite.
Provided that the condition (\ref{23}) is verified, one can show that the
gswf's eq. (\ref{24}) give a complete set of states  with
the correct boundary conditions in each particle variable$^{16}$.

There is a strict correspondence between  modular invariance of
2DCFT and  gauge invariance of the physical system.
Indeed, by using the usual modular transformations for the
$\theta_1$-functions$^{15}$
and appropriate transformation properties
for the center of charge
theta functions
under the transformation $\tau
\rightarrow \tilde \tau =-\frac 1\tau $, and
$ z\rightarrow \widetilde{z}=\frac z\tau $, namely
\bey
\Theta \left[
\begin{array}{c}
K^{-1}\vec l \\
\vec 0
\end{array}
\right] \left( K\tau ^{-1}\vec Z|-K\tau ^{-1}\right)= \det \left( \frac{%
K^{-1}\tau }i\right) ^{1/2}\exp \left[ \pi i\vec Z_TK\tau ^{-1}\vec Z\right]
\times
\nonumber
\\
\times
\sum_{\vec l^{\prime }\in \text{{\bf Z}}_K^n}\exp \left[ -\pi
i\vec l_TK^{-1}\vec l^{\prime }\right] \Theta \left[
\begin{array}{c}
K^{-1}\vec l \\
\vec 0
\end{array}
\right] \left( K\vec Z|K\tau \right),
\eey
one sees that, provided the condition (\ref{23}) is satisfied, the following
equation holds
\begin{equation}
\label{27}
F_{\vec l}\left( \left\{ \tilde z_i^I\right\} |\tilde \tau
\right) =\frac c{\sqrt{\det K}}\exp \left[ \frac{\pi i}\tau N_\Phi
\sum_{I=1}^n\sum_{i=1}^{N_I}\left( z_i^I\right) ^2\right] \sum_{\vec l^{\prime
}\in \text{{\bf Z}}_K^n}\exp \left[ -2\pi i\vec l_TK^{-1}\vec l^{\prime
}\right] F_{\vec l^{\prime }}\left( \left\{ z_i^I\right\} |\tau \right),
\end{equation}
where $c$ is an inessential constant.
The physical meaning of the above equation is more easily understood when
we complete the gswf $\Psi _{\vec l}$ by multiplying its analytic part eq.
(\ref{24}) by the appropriate gaussian factor, obtaining
\begin{equation}
\label{28}
\Psi _{\vec l}\rightarrow \tilde \Psi _{\vec l}
=\frac {c'}{\sqrt{\det K}}\exp \left[ 2\pi iN_\Phi
\sum\limits_{I=1}^n\sum\limits_{i=1}^{N_I}\left( \xi _i^I\eta _i^I\right)
\right] \sum\limits_{\vec l^{\prime }\in \text{{\bf Z}}_K^n}\exp \left[
-2\pi i\vec l_TK^{-1}\vec l^{\prime }\right] \Psi _{\vec l^{\prime }}
\end{equation}
As the exponential
factor correspond to the gauge transformation taking from the gauge
$ \vec A _1=2\pi N_\Phi \left( \eta ,0\right) $
to the gauge
$ \vec A _2=2\pi N _{\Phi} \left( 0,- \xi \right) $,
we see that the eq. (\ref{28}) express the gauge invariance
of the theory.

The global properties of the gswf's are encoded in the center of charge
theta function. Let us introduce the total magnetic translation group
\begin{equation}
\label{29}
\begin{array}{l}
{\cal S}_b  F_{\vec l} \, \left(\left\{ z^I_i \right\} \right)=
F_{\vec l} \, \left( \left\{ z^I_i+b \right\} \right),
\\
{\cal T}_a F_{\vec l} \, \left(\left\{ z^I_i \right\} \right)=
\exp \left[ 2\pi iN_\Phi
\left(\frac aL\sum_{I=1}^nZ^I+\tau \frac{a^2}{2L^2}\sum_{I=1}^nN_I\right)
\right] F_{\vec l} \, \left( \left\{ z^I_i +a\tau \right\} \right) \, .
\end{array}
\end{equation}
${\cal S}_b$ e ${\cal T}_a$ are non commuting operators, in particular by
taking
$a=b=L/N_\Phi $, one has
\begin{equation}
\label{31}
{\cal S}_{\frac{L}{N_\Phi }}{\cal T}_{\frac{L}{N_\Phi }%
}=\exp \left[ 2\pi i\sum_{I,J=1}^n\left( K^{-1}\right) ^{IJ}\right] {\cal T}%
_{\frac{L}{N_\Phi }}{\cal S}_{\frac{L}{N_\Phi }}=
e^{2\pi i\nu} {\cal T}
_{\frac{L}{N_\Phi }}{\cal S}_{\frac{L}{N_\Phi }}\, .
\end{equation}
The gswf's
$\left\{ F_{\vec l}\, , \,\vec l\in \text{{\bf Z}}_K^n\right\} $
provide
a basis
for an irreducible representation of
the subgroup of the magnetic
translation group generated by
${\cal S}_{L/N_\Phi }$ e ${\cal T%
}_{L/N_\Phi }$.
In the language of 2DCFT
such operators are explicitly realized in terms of holomorphic field
$\vec \phi \left( z\right) $
as follows
\begin{equation}
\label{32}
\begin{array}{l}
{\cal S}_{\frac{L}{N_\Phi }}=\exp \left[ \vec 1_T R^{-1}\oint_Adz\partial
\vec \phi \left( z\right) \right]\, ,
\\
{\cal T}_{\frac{L}{N_\Phi }}=\exp \left[ \vec 1_T R^{-1}\oint_Bdz
\partial \vec \phi \left( z\right) \right] \, ,
\end{array}
\end{equation}
where $\vec 1_T=\left( 1,...,1\right) $,
and the integrals are taken along the homology cycles of the torus.
The translation operator along the $B$ cycle enters directly the evaluation
of the conductance of the system. As for the Laughlin states,
we introduce a magnetic flux line
of strength
$\Delta \Phi =2\pi \lambda $
along the $B$ cycle of the torus.
The corresponding change in the periodicity of the gswf's along the $A$ cycle
implies the transformation
\begin{equation}
\label{33}
F_{\vec l}\rightarrow F_{\vec l+\lambda {\vec 1}} \, ,
\end{equation}
that can be explicitly realized by means of the action
of the operator
${\cal T}_{\lambda L/N_\Phi }$ on  $F_{\vec l}$ \, :
\begin{equation}
\label{34}
F_{\vec l+\lambda { \vec 1 }}\left( \left\{ z_i^I\right\}
|\tau\right) ={\cal T}_{\frac{\lambda L}{N_\Phi }}F_{\vec l}\left( \left\{
z_i^I\right\} |\tau\right) =\left[ \prod\nolimits^{\prime} \theta _1\right]
\Theta \left[
\begin{array}{c}
K^{-1}\left( \vec l+\lambda
{ \vec 1}\right) \\ 0
\end{array}
\right] \left( K\vec Z|K\tau \right)\, .
\end{equation}
As the electric field corresponding to the change of flux is along the A cycle
and
the translation is along
the $B$ cycle, the conductance is purely transverse and is given by Faraday
law:
\begin{equation}
\label{35}
\sigma _H=\frac Q{\Delta \Phi}=\sum_{I,J=1}^n\left(
K^{-1}\right) _{IJ} =\frac {\nu}{2 \pi}  \, .
\end{equation}

{\bf 4. Edge states for the generic abelian case}

As in the simple case of the Laughlin states $\nu=1/m$, the
dynamics of the edge states is, for the generic abelian case, completely
dictated by the same 2DCFT leading to the bulk gswf's.
As a consequence the hamiltonian for the edge
excitations is simply given as a superposition
of the free hamiltonian relative to each component of the chiral field
$\vec \phi$ propagating on the edge of the sample,
$\phi _I=\phi _I (x-v_It)$:
\begin{equation}
\label{36}
{\cal H}= \sum_{I=1}^n {\cal H}^I=\sum_{I=1}^n \frac {v_I}{8\pi }
\left[ \frac 1{v_I^2}\left( \partial
_0\phi _I\right) ^2+\left( \partial _x\phi _I\right) ^2\right]
\end{equation}
where $v_I$ is the propagation velocity of the $I^{th}$ mode.
The electromagnetic field does not couples with the same strength to
each component of the field $\vec \phi$. In order to find the correct
structure of the electromagnetic current we require that it should lead to
the physical value of the Hall conductance. We can therefore read its
structure out of the form of the magnetic translation operator eq. (\ref{32}).
Then
\be
\label{J}
J=\sum_{I=1}^n j^I \, ,
\ee
where
\begin{equation}
\label{38}
j^I=\sum_{J=1}^n \frac 1{2\pi }  {\left( R^{-1} \right)}_{IJ}
\partial _x\phi _J\, .
\end{equation}
Notice that currents satisfy the ${\hat {U}(1)}^n$  Kac-Moody algebra,
that in terms of
Fourier components reads
\begin{equation}
\label{39}
\left[ j _k^I,j _{k^{\prime }}^J\right] =\left(
K^{-1}\right) _{IJ}k\delta _{k+k^{\prime },0} \, ,
\end{equation}
in agreement with the results of ref. 10.

The hamiltonian eq. (\ref{36}) can be rewritten in
terms of the physical edge currents instead of the chiral field
$\vec \phi $:
\begin{equation}
\label{40}H=2\pi \sum_{I,J=1}^n\sum_{k>0}V^{IJ}j_k^I j_{-k}^J ,
\end{equation}
where
\begin{equation}
\label{41}
\left( R_T^{-1}VR^{-1}\right) _{IJ}=v_I\delta _{IJ} .
\end{equation}
Notice that a non diagonal compactification matrix implies an interaction
between edge currents.

The spectrum of all possible charged edge excitations is again given by the set
of $\det K$ VO's, eq. (\ref{22}), written in terms of the chiral field
$\vec \phi$
defined on the edge, and the corresponding charge is obtained trough
the equation
\be
\label{comm}
\left[ J\left( x\right) ,V_{\vec l}\left( x^{\prime }\right)
\right] =\delta \left( x-x^{\prime }\right) \sum_I\left( K^{-1}\right)
_{IJ}l_J V_{\vec l}\left( x\right) \, .
\ee
We see that the normalization
of the currents chosen in  eqs. \ref{J}, (\ref{38}) in order to be
consistent with the Hall conductance determines the correct value of the
charge as is given by eq. (\ref{q}).
The introduction of the $n$ component vector $\vec \phi$ leads then for
a generic abelian Hall fluid to the existence of $n$ branches of charged
excitations, each corresponding to a $\hat {U}(1)$ current. We shall see in
next section that for a certain set of $K$ matrices a different structure
for 2DCFT arises. As a consequence, the
structure of the edge states will be correspondingly modified.

{\bf 5. Jain's sequences}

The description of a generic abelian Hall
fluid, as a $n$ component 2DCFT
characterized by an integer valued,
non singular symmetric matrix $K$,
is essentially kinematical in nature, and does not lead to specific dynamical
prediction on the stability of the different states. In contrast, there is a
striking experimental evidence that
the most prominent Hall
plateaux are seen at the fillings of the principal sequence
$\nu=n/(2n \pm 1)$ and of the next stable sequence at $\nu=n/(4n \pm 1)$.
It was first suggested by Jain$^{13}$ the idea of looking at the FQHE for
electrons at filling $ n/(2pn \pm 1) $ as
a manifestation of the integer effect for {\it composite
fermions}, obtained by attaching to each electron an even
number of flux units opposite to
the external magnetic field. This approach has found further evidence in
observation that
the energy gaps, measured for the principal  sequence$^{17}$, correspond to the
cyclotron energies relative to the reduced magnetic field $B-B_{1/2}$
and from  the analysis
of the accumulation point of the principal sequence at $ \nu = 1/2 $, showing
the existence of many features typical of a Fermi surface$^{18}$.
Furthermore, in this framework there is a natural
explanation for measurements of non-local four terminal
magneto-resistance$^{19}$,
showing that there is no indication
of edge state dissipationless conduction near filling $\nu=1/2$.
However it has been shown that Jain's approach can be extended
to an arbitrary abelian Hall fluid by introducing a non trivial connection
that takes into account the braiding factor between any two
particles$^{20}$. All efforts to clarify the specific nature of
Jain's states is, therefore, extremely relevant.

Let us then recall that for the Jain's sequences corresponding to filling
factors $\nu =n/ (2pn+1)$
the structure of the $K$ matrix is the following:
\be
\label{44}
K = \text{{\bf 1}}+2p C \, ,
\ee
where $\text {{\bf 1}}$  is the unit $n \times n$ matrix and $C$ is an
$n \times n$ matrix with all entries equal to $1$.
The corresponding compactification matrix is given by
\be
\label{47}
R =\text {{\bf 1}} +{1\over n}(\sqrt{n\over \nu} -1)C \, .
\ee
Notice that  $R$ is transformed in its inverse by sending $\nu$ into  the
integer value $\nu '=n^2/\nu$.
By recalling that $\vec{\b}_I = R \vec{\d}_I$ we see that
\be
\label{48}
\vec \b_I -\vec \b_J =\vec \d_I -\vec \d_J \, .
\ee

We shall call a matrix $K$, such that the difference between two of the
corresponding vectors $\vec{\b}_I$ has integer entries, {\it degenerate}. The
rational behind this denomination is that, as it has been shown by Cappelli,
Trugenberger and Zemba$^{14}$, the corresponding Hall fluid is characterized by
a
degenerate representation of the $W_{1+\infty}$ algebra.

In order to unveil the consequence of the peculiar structure of the $K$
matrix given by the eq. (\ref{44}), we recall that it can be diagonalized
by means of an orthogonal
transformation, $K=O_T K_{diag} O$, where
\be
\label{46}
O= \left(
\begin{array}{ccccccc}
1/\sqrt{2} & -1/\sqrt{2} & 0 & ... & ... & ... & 0 \\
1/\sqrt{6} & 1/\sqrt{6} & -2/\sqrt{6} & 0 & ... & ... & 0 \\
1/\sqrt{12} & 1/\sqrt{12} & 1/\sqrt{12} & -4/\sqrt{12} & 0 & ... & 0 \\
... & ... & ... & ... & ... & ... & 0 \\
1/\sqrt{n} & 1/\sqrt{n} & ... & ... & ... & ... & 1/\sqrt{n}
\end{array}
\right)
\ee
We introduce the new set of fields
\bey
\Phi_i &=&O_{ij} \phi_{j}, \quad i= 1,2,\dots n-1 \, ,
\\
\Phi_+ &=&O_{nj}\phi_j  ={1\over \sqrt {n}}(\phi_1 + \phi_2 \dots \phi_n)\, .
\eey
As the matrix $O$ is orthogonal the new fields are still independent:
\be
\label{50}
\left\langle\Phi_i(z)\Phi_j(z')\right\rangle =-\d_{ij} \ln (z-z') \, .
\ee
We can then write the scalar product $\vec{\b}_I\cdot \vec{\phi}(z)$ in the
form
\be
\label{51}
\vec{\b}_I\cdot \vec{\phi}(z) ={1 \over {\sqrt\nu}}
\Phi_+(z)+\vec{u}_J\cdot\vec{\Phi}(z) \, ,
\ee
where the $n$, $(n-1)$-dimensional vectors $\vec{u}_I$ are given by
the columns of the matrix $O$ with the last row omitted:
\be
\label{52}
O=\left(
\begin{array}{cccc}
{\vec u}_1 & {\vec u}_2 & ... & {\vec u}_n \\
1/\sqrt{n} & 1/\sqrt{n} & ... & 1/\sqrt{n}
\end{array}
\right).
\ee
They satisfy the following relations
\be
\label{53}
\sum_{I=1}^n{\vec u}_I=0,
\quad \sum_{I=1}^n ({\vec u}_I)_a({\vec u}_I)_b=
\delta _{ab},
\quad {\vec u}_I \cdot {\vec  u}_J=
\delta _{IJ}-\frac 1n,
\ee
and are strictly related to the lattice of roots and weights of $SU(n)$:
the simple roots $\vec{\a}_I$ are explicitly given by
\be
\label{54}
{\vec \alpha}_I={\vec u}_I -{\vec u}_{I+1},
\ee
and the fundamental weight $\vec{\L}_I$ by
\be
\label{55}
{\vec \Lambda}_I =\sum_{J=1}^I {\vec u}_J,\quad
I=1,...,n-1\,.
\ee
Out of the VO's $V_{{\vec u}_I}=:\exp (i\, \vec{u}_I\cdot
\vec{\Phi}):$
one can build the currents corresponding to the off-diagonal generators of
the affine $\hat {SU}(n)$ algebra$^{21}$
\be
\label{56}
J_{\vec{\a}} = :e^{i\vec{u}_I\cdot\vec{\Phi}}
e^{
 -i\vec{u}_{J}\cdot\vec{\Phi}}:\, ,
\ee
where $\vec \a ={\vec u}_I-{\vec u}_J$ is a root of $SU(n)$.
In order to close the $\hat {SU}(n)$ affine algebra, one has to introduce the
diagonal currents
\be
\label{57}
J_I=:e^{i\vec{u}_I\cdot\vec{\Phi}}
e^{-i\vec{u}_{I}\cdot\vec{\Phi}}:=i \pde\Phi_I , \quad I=1,\dots,n-1.
\ee
The group structure can be easily seen by looking at the operator
product
expansion (OPE) for the currents:
\bey
J_I(z)J_J(w) & \sim & 0\, ,
\\
J_{\vec \a}(z)J_{-\vec \a}(w) & \sim & {1 \over (z-w)^2}+{ \sum_{I=1}^{n-1}
(\vec \a)_I
J_I(w)\over (z-w)}\, ,
\\
J_I(z)J_{\vec \a}(w) & \sim & -{( \vec \a)_I
J_{\vec\a}(w)\over (z-w)}\, ,
\\
J_{\vec \a}(z)J_{\vec \a'}(w) & \sim & {
J_{\vec \a +\vec \a'}(w) \over (z-w)}\, ,
\eey
where the last equation holds if $\vec \a +\vec \a'$ belongs to the root
lattice, otherwise the right hand side is zero.

It is interesting to notice that only the field ${\Phi}_+$ contributes to the
Hall conductance, implying that it is the only charged mode. This can be seen
by recalling that the conductance is obtained through the action of the
magnetic translation operator along the $B$ cycle of the torus, eq. (\ref{34}),
and noticing that
\be
\label{58}
{\cal T}_{\frac{L_1}{N_\phi }}=\exp
\left[ \vec 1_tR^{-1}\oint_Bdz\partial \vec \phi \left( z\right) \right]=
\exp
\left[ \sqrt{\nu}\oint_Bdz\partial {\Phi}_+ \left( z\right) \right]
\end{equation}

This suggests of looking  in some detail
the structure of the center of charge theta functions responsible for the
global properties of the system.
We first show that they will decompose into a sum of terms, that are products
of the contribution of the charge and neutral modes.
To this  purpose, let us introduce the matrix $\L$
given by $\L=DO$ where $O$ is given by
eq. (\ref{46}) and $D$ is a diagonal matrix with the following entries
\be
\label{60}
D=\text{{diag}} \left( \sqrt{2},\sqrt{6},...,\sqrt{i\left( i+1\right) },...,
\sqrt{\left( n-1\right) n},\sqrt{n}\right).
\ee
Its usefulness can be seen by noticing that it diagonalizes both $K$
and $K^{-1}$, $\L K=K_{diag}\L$,
$\L K^{-1}=K_{diag}^{-1}\L$,
that the new variables $\vec W=\L \vec Z$ are such that only
$W_n=Z_1+\dots+Z_n$ is affected by total translations, while the
remaining are left invariant, and finally that the center of charge theta
function can be written as a theta function on the lattice $\L
{\text{{\bf Z}}}^n$, namely:
\be
\label{theta}
\Theta \left[
\begin{array}{c}
K^{-1}\vec l \\
\vec 0
\end{array}
\right] \left( K\vec Z|K\tau \right) =\Theta _\Lambda \left[
\begin{array}{c}
K_{diag}^{-1}\vec L \\
\vec 0
\end{array}
\right] \left(  K_{diag}D^{-2}\vec W| K _{diag}D^{-2}\tau \right)
\ee
where $\vec L=\Lambda \vec l$, and the function $\Theta _{\L}$ is defined as
in eq. (\ref{25}), except that the sum over $\vec h$ runs on $\vec h
 \in \L
{\text {{\bf Z}}}^n$ instead of $\vec h \in {\text{{\bf Z}}}^n$.
The vectors $\vec L$, that  belong to $\Lambda \text{{\bf Z}}^n$,
are defined modulo $\Lambda K\text{{\bf Z}}^n=K_{diag}\Lambda \text{{\bf Z}}^n$
and one can show that a set of inequivalent values is given by
$\left\{\vec L=ln\vec \delta _n;l=1,\dots,2pn+1\right\}$, corresponding to
$\left\{\vec l=l\vec 1;l=1,\dots,2pn+1\right\}$. Since
the lattice $\Lambda \text{{\bf Z}}^n$ can be expressed
as follows:
\begin{equation}
\L \text{{\bf Z}}^n=
\bigcup_{a=1}^{\det \Lambda}
\left\{ \vec r^{\,\left( a \right) }+D^2\text{{\bf Z}}
^n\right\}\,
\end{equation}
where $\det \Lambda=n!$,
and the explicit expression for the integer
vectors $\vec r^{\,(a)}$ is easily evaluated in any specific case,
we obtain the following decomposition of the center of charge theta
function:
\bey
\label{63}
\Theta \left[
\begin{array}{c}
\frac l{2pn+1} \vec 1 \\
\vec 0
\end{array}
\right] \left( K\vec Z|K\tau \right)& =&
\sum_{a=1}^{\det \Lambda} \left\{ \prod_{i=1}^{n-1}\Theta \left[
\begin{array}{c}
\frac{ r_i^{\left( a \right) } }{i\left( i+1\right) }
\\ 0
\end{array}
\right] \left( W_i|i\left( i+1\right) \tau \right)\right. \times
\nonumber
\\
  & \times &
\left. \Theta \left[
\begin{array}{c}
\frac{r_n^{\left( a \right) }}n+\frac{\nu l}{n} \\ 0
\end{array}
\right] \left( \frac n\nu W_n|\frac{n^2}\nu \tau \right) \right\}.
\eey
It is clear that in each term of the sum
only the factor depending on $W_n$ contributes to the Hall conductance
and corresponds then to the charged mode, while the remaining $n-1$
correspond to the neutral ones.

In order to stress the connection of this decomposition
with the structure $\hat {U}(1)\otimes \hat {SU}(n)$ that we have
discussed with reference to VO's, we take $\vec Z=0$, recovering
the characters of our 2DCFT. For
the sake of simplicity, we will analyze the case $n=2$ and $n=3$.
In the first case one has:
\bey
\label{64}
\Theta \left[
\begin{array}{c}
\frac l{4p+1}\vec 1 \\
\vec 0
\end{array}
\right](\vec 0|K\tau)&=&
\Theta \left[
\begin{array}{c}
0 \\
0
\end{array}
\right] \left(0|2\tau \right) \Theta \left[
\begin{array}{c}
\frac 12 +\frac l{4p+1} \\
0
\end{array}
\right] \left(0|4\tau/\nu \right)
+
\nonumber
\\
&+&\Theta \left[
\begin{array}{c}
1/2 \\
0
\end{array}
\right] \left(0|2\tau \right)
\Theta \left[
\begin{array}{c}
\frac l{4p+1}\\
0
\end{array}
\right] \left(0|4\tau/\nu \right)\, ,
\eey
that correspond to the character decomposition
\be
\label{char}
\chi^l=
\chi_{SU(2)_1,\L =0}
\chi_{U(1)}^{l,1}
+\chi_{SU(2)_1,\L=1/2}
\chi_{U(1)}^{l,0}
\ee
where $\chi_{SU(2)_1,s}$ is the character of the affine $\hat {SU}(2)$ level 1
representation of highest weight $s$ and $\chi_{U(1)}^{l,i}$ is the
the character of a $\hat {U}(1)$ theory of conformal weight $1/ 2\nu$,
summed over the lattice $2\text{{\bf Z}}+i+l\nu$,
where $l$ (resp. $i$) is defined
modulo $4p+1$ (resp. $2$).
In a similar
fashion for $n=3$ we have:
\be
\label{65}
\chi^l=\chi_{SU(3)_1,\vec\L=(0,0)} \chi_{U(1)}^{l,0}
+\chi_{SU(3)_1,\vec\L=(1/\sqrt 2,1/\sqrt 6)}
\chi_{U(1)}^{l,1}
+ \chi_{SU(3)_1,\vec\L=(0,1/\sqrt 6)}
\chi_{U(1)}^{l,2}
\ee
where the notation relative to $\hat {SU}(3)$ characters is self explanatory
and $\chi_{U(1)}^{l,i}$ is the character of a $\hat {U}(1)$ theory
of conformal
weight $1/2\nu$
summed over the lattice $3\text{{\bf Z}}+i+ l\nu $, with $l$ modulo
$6p+1$ and $i$ modulo $3$.

Finally, by eq. \ref{36}, it immediately seen that the extended algebra
$\hat{U}(1)\otimes \hat{SU}(n)$ introduced in terms of the VO's correspond
to a symmetry of the edge dynamics as long as the difference in
velocity among the components is disregarded.
Furthermore   the structure of the edge excitations is again determined by the
general forms of chiral the VO's eq. (\ref{22}),
on
the edge. In terms of the new basis the generic VO is given by
\be
V_{\vec l}(z)=:e^{i\left({{\sqrt \nu} \over n}\sum_{I=1}^{n-1}l_I\right)\Phi
_+}:
:e^{i\sum_{I=1}^{n-1}l_I
{\vec u}_I\cdot {\vec \Phi}}: \,
, \quad \quad \quad \vec l\in \frac{%
\text{{\bf Z}}^n}{K\text{{\bf Z}}^n}\equiv \text{{\bf Z}}_K^n \,
\ee
Notice that the electric current, eq. \ref{J}, is written only in terms of
the $\hat {U}(1) $ mode:
\be
\label{J+}
J={1 \over 2 \pi} {\sqrt \nu}\pde _x \Phi _+
\ee
and has the same structure as in the Laughlin case. Therefore, the more
general excitation correspond to the propagation of a single charged mode
and $n-1$ neutral ones. It is also suggestive to notice that
for the choice of inequivalent vectors $\vec l=l \vec 1$
with $l=1,2, \dots, \det K=2np+1$
the set of independent VO's takes the form
\be
\label{VJL}
V_{ l}(z)=:e^{il{\sqrt \nu} \Phi _+}: \,
, \quad \quad \quad  l\in
\frac{\text{{\bf Z}}}{(2np+1)\text{{\bf Z}}} \, ,
\ee
in complete analogy to the case of filling $\nu=1/m$, eq. \ref{4}.

{\bf 6. Concluding remarks}

In this paper we have shown how construct a consistent 2DCFT
description of an arbitrary abelian
Hall fluid, under the requirement that the corresponding $K$ matrix is
positive definite. Under this
condition, that express the basic requirements of having
a consistent description on higher genus Riemann surfaces,
one obtains a complete characterization of all universal properties, such as
the Hall conductance and the spectrum of low energy excitations, that are
experimentally accessible as edge state excitations of the system.
This theory is kinematical in nature and therefore is not expected to
give any dynamical information on the relative stability of the different
Hall fluids. However it already leads to a
natural distinction between the case of a generic abelian fluid  and the case
of a fluid  belonging to a Jain's sequence, where  an extended
$\hat{U}(1) \otimes \hat{SU}(n)$ algebra appears.
In view of the experimental evidence showing that the most prominent Hall
plateaux correspond to Jain's sequences, it is then natural to ask whether
this difference  may play
a dynamical role in selecting the most stable
abelian quantum Hall fluids. At an intuitive level the analysis of sec. {\bf
5} seems to indicate that this should be the case. In fact any excitation of
a Hall fluid belonging to a Jain's sequence may be described, as shown by
eq. (\ref{VJL}), in complete analogy to a single component Laughlin's state,
regardless of the number $n$ of components.

Looking at this problem more closely, it is interesting to notice that
the conformal properties of the VO's corresponding to tunneling
transitions between
different channels determine the relevance of the corresponding impurity
driven transition amplitudes, at least at a perturbative level. Therefore, the
$K$ matrix contains at a perturbative level full information on the role of
disorder for an abelian  Hall fluid. What is then typical of the Jain's
states$^{12}$,
under the positivity condition of the $K$ matrix,
is that the renormalization group (RG)
behaviour for the disorder driven tunneling amplitudes is {\it universal},
that is independent on the values of $p$ and $n$ and of the edge velocities.
This leads to a new fixed point quadratic action, that takes completely
into accounts disorder and is insensitive to interactions between different
channels, leaving the charged mode uncoupled from the neutral ones.

The
situation in completely different for a generic abelian Hall fluid,
for which all modes are charged and  the
RG  behaviour of the transition amplitudes is determined by the detailed
structure of the $K$ matrix. This may leads, for a given system,
to transition amplitudes that are relevant or irrelevant
according to the choice of different couples of channels.
This suggest that the study of the role of the disorder for a generic abelian
Hall fluid may be useful for understanding the experimental observed
stability of different fillings.

{\bf Acknowledgements}

Discussions with G. Cristofano, G. Maiella, F. Nicodemi and P. Sodano
are gratefully acknowledged.

Work supported in part by MURST and by EEC contract n. SC1-CT92-0789.

\begin{center}
{\bf References}
\end{center}

\begin{enumerate}

\item D.C. Tsui, H. L. Stormer and A. C. Gossard, Phys. Rev. Lett. {\bf
48}, 1559 (1982).

\item See e.g. M. Stone (ed.),
{\it Quantum Hall Effect}, World Scientific Singapore (1992).

\item M. B\"uttiker, Phys. Rev. Lett. {\bf 57}, 1761 (1986).

\item  For a review see e.g. G. Cristofano, G. Maiella, R. Musto and F.
Nicodemi,
Nucl. Phys. {\bf 33C}, 119 (1993).

\item R. B. Laughlin, Phys. Rev. Lett. {\bf 50} 1395 (1983)

\item A. De Martino and R. Musto,
 {\it Knizhnik-Zamolodchikov equation and extended symmetry for stable
Hall states }, preprint DPS-NA-19
/95, cond-mat 1995.

\item X. G. Wen, Phys. Rev. {\bf B 41}, 12838 (1990).

\item G. Cristofano, G. Maiella, R. Musto and F. Nicodemi, Mod. Phys. Lett.
{\bf 6A} 2885 (1991).

\item  See e.g. J. Fr\"olich and A. Zee, Nucl. Phys. {\bf B364}, 517 (1991);
X. G. Wen and A. Zee, Phys. Rev. {\bf B 46}, 2290 (1993).

\item X. G. Wen, Int. J. Mod. Phys. {\bf B6}, 1711 (1992), and references
contained therein.

\item C. L. Kane, M. P. A. Fisher and J. Polchinski, Phys. Rev. Lett. {\bf 72},
4129 (1994).

\item  C. L. Kane and M. P. A. Fisher, {\it Impurity scattering and transport
of
fractional Quantum Hall edge states}, preprint cond-mat 9409028.

\item  J. K. Jain, Phys. Rev. Lett. {\bf  65}, 199 (1989); Phys. Rev. {\bf B
40}, 8079 (1989); {\bf B 41} 7653 (1990).

\item  A. Cappelli, C. A. Trugenberger and G. R. Zemba, {\it Stable
Hierarchical Quantum Hall Fluids as W$_{1+\infty }$ Minimal Models}, preprint
UGVA-DPT 1995/01-879, DFTT 09/95, hep-th 9502021.

\item D. Mumford, {\it Tata Lectures on Theta}, Birkh\"auser Boston (1983).

\item E. Keski-Vakkuri and X. G.
Wen, Int. J. Mod. Phys. {\bf B7}, 4227
(1993).

\item   R. R. Du, H. L. Stormer, D. C. Tsui, L. N. Pfeiffer  and  K. W. West,
Phys. Rev. Lett. {\bf 70}, 2944 (1993).

\item  See e.g. B. I. Halperin, P. A . Lee and N. Read, Phys. Rev. {\bf B
47}, 7312 (1993); S. H. Simon and B. I. Halperin, Phys. Rev. {\bf B 48} 17368
(1993),
{\bf B 50} 1807 (1994);

\item  J. K. Wang and V. J. Goldman, Phys. Rev. Lett. {\bf 67}, 74 (1991)

\item  G. Cristofano, G. Maiella, R. Musto and F. Nicodemi, {\it FQHE and
Jain's approach on the torus}, Int. J. Mod. Phys. {\bf B}, in press;
M. Flohr and R. Varnhagen, Jour. Phys. {\bf A27}, 3999 (1994);
D. Karabali, Nucl. Phys. {\bf FS B419}, 437 (1994).

\item  see e.g. P. Goddard and D. Olive, Int. J. Mod. Phys. {\bf A1}, 303
(1986).

\end{enumerate}

\end{document}